\documentclass[letter]{aa}
\usepackage{hyperref}
\usepackage{comment}
\usepackage[flushleft]{threeparttable}
\usepackage[dvipsnames]{xcolor}
\usepackage{soul}
\usepackage{mathtools}
\usepackage{caption}
\usepackage{dblfloatfix}
\usepackage{placeins}
\usepackage{array}

\usepackage{xcolor}
\hypersetup{    
    colorlinks=true,
    citecolor = blue,
    linkcolor=blue,
    filecolor=magenta,      
    urlcolor=cyan,
}
\usepackage{multirow}
\usepackage{multicol}
\usepackage{pdflscape} 
\usepackage{graphicx}
\usepackage{txfonts}

 \def\mso{\,\mathrm{M_\odot}}

 \def\Pi{P_\mathrm{i}}

 \def\gr{(g-r)_{r_\mathrm{max}}}
 \def\gap{\Delta {(g-r)_{r_\mathrm{max}}}}
\newcommand{\Fig}[1]{Fig.\,\ref{#1}}
\newcommand{\Figure}[1]{Figure\,\ref{#1}}

\newcommand*{\rev}[1]{\textcolor{black} {#1}}

\begin{document} 

   \title {Type Ib supernovae are bluer than Type Ic supernovae}

   \author{Harim Jin\inst{1}, 
           Selma E. de Mink\inst{1},
           Sebastian Holzner\inst{2,3},
           Jakub Klencki\inst{1}, 
           G\'eza Csörnyei\inst{2}, 
           Sung-Chul Yoon\inst{4},  
           Iair Arcavi\inst{5}, 
           Wolfgang E. Kerzendorf\inst{6}
          }

   \institute{Max Planck Institute for Astrophysics, Karl-Schwarzschild-Straße 1, 85748 Garching bei München, Germany\\
              \email{jin@mpa-garching.mpg.de}
    \and
    European Southern Observatory, Karl-Schwarzschild-Str. 2, 85741 Garching, Germany
    \and
    Technische Universit\"at M\"unchen, TUM School of Natural Sciences, Physics Department, James-Franck-Stra\ss{}e 1, 85748 Garching, Germany
    \and
    Department of Physics and Astronomy, Seoul National University
    \and
    The School of Physics and Astronomy, Tel Aviv University, Tel Aviv, 69978, Israel
    \and
    Department of Physics and Astronomy, Michigan State University, East Lansing, MI 48824, USA}

   \date{Received Month Day, XXX; accepted Month Day, XXX}

  \abstract
   {Type Ib and Ic supernovae (SNe Ib/Ic) are the bright finale of massive stars that have lost their hydrogen envelopes, making them powerful probes of mass stripping in massive star evolution. The advent of modern large photometric and spectroscopic surveys presents a unique opportunity to investigate systematic differences between these two kinds of SNe. In this study, we analyze a large, homogeneous sample of SNe Ib/Ic light curves from the Zwicky Transient Facility. We find a systematic difference in their \rev{apparent} optical colors \rev{at peak}: SNe Ib are, on average, bluer than SNe Ic\rev{, with two-sample tests confirming that the distributions differ ($p<0.05$). The difference in their host galaxy reddening, as currently constrained, is too small to fully explain it.} 
   \rev{The color} difference \rev{therefore most likely has an intrinsic origin}, reflecting progenitors with different degrees of stripping -- helium-rich for SNe Ib and helium-poor for SNe Ic. In addition, we find that SNe Ib/Ic with narrow lines (SNe Ibn/Icn) are bluer than those without, which might originate from circumstellar matter interaction, with potential connection to fast blue optical transients. We demonstrate that SN colors offer a promising probe of mass stripping in massive stars, potentially providing a useful tool for analyzing large photometric data and improving predictions for the final outcomes of stripped massive stars.
}


   \maketitle

\section{Introduction}\label{sec_intro}

Type Ib and Ic supernovae (SNe Ib/Ic) are explosions of stripped massive stars and provide valuable insights into how these stars lose their outer layers through stellar winds, outbursts, or binary interactions \citep[e.g.,][]{Podsiadlowski1992,Woosley2002,Meynet2005,Yoon2010,Eldridge2013}. Most known stripped stars, such as Wolf-Rayet stars or intermediate-mass stripped stars, are still in their core helium burning phase \citep[e.g.,][]{Crowther2007,Drout2023}. 
However, their subsequent evolution remains uncertain: 
how much further stripped they become before core collapse, and whether they undergo an additional mass transfer shortly before SN, remain open questions \citep{Laplace2020,Klencki2022,Goetberg2023}.
Addressing these questions are essential for understanding the chemical, dynamical, and radiative feedback of stripped stars \citep[e.g.,][]{Izzard2006,Stanway2016,Goetberg2020}, the formation of X-ray binaries \citep{Valli2025}, the origin of gravitational-wave sources \citep[e.g.,][]{Kruckow2018,Laplace2020}, and the role of binary interactions in shaping the circumstellar medium (CSM) in which SNe explode \citep[e.g.,][]{Ercolino2025}. Direct detection of stripped stars in later evolutionary phases, or shortly before the explosion, is expected to be extremely challenging \citep[e.g.,][]{Yoon2012}. As a result, the late evolution of stripped massive stars remains poorly constrained, and SNe Ib/Ic provide a crucial -- perhaps unique -- insight into these otherwise elusive late evolutionary histories.

This is a great time to study SNe Ib/Ic to investigate the lives of stripped massive stars. Statistically significant samples of SNe Ib (showing helium features in their optical spectra) and Ic \citep[lacking \rev{clear optical} helium features;][]{Filippenko1997,Modjaz2014,GalYam2017} are now available thanks to the advent of large-scale automated transient surveys such as the Zwicky Transient Facility \rev{\citep[ZTF;][]{Bellm2019,Fremling2020,Perley2020}}, allowing for population-level studies \citep[e.g.,][]
{Taddia2015,Prentice2019,Afsariardchi2021,Barbarino2021,Rodriguez2023}. While up to 2010 only a few hundred SNe Ib/Ic were known, today about a hundred new SNe Ib/Ic are being discovered each year\footnote{Transient Name Server (TNS) \url{https://www.wis-tns.org/}}, which will \rev{only increase in the upcoming Legacy Survey of Space and Time (LSST) era \citep{Ivezic2019}.}
The large number of SNe Ic uncovered by these surveys is in apparent tension with standard binary evolution models, which predict non-negligible helium masses \rev{retained at} core collapse. This tension has led to an ongoing debate about whether additional helium stripping is required or helium could be spectroscopically hidden \citep[e.g.,][]{Hachinger2012,Yoon2015,Dessart2020,Lu2026} 

The colors of SNe Ib/Ic at the optical peak offer an independent method of probing the amount of helium to tackle this debate and the nature of SNe Ib/Ic progenitors \citep{Jin2023} (hereafter JYB23).
Building on JYB23, we study the peak color difference \rev{between SNe Ib and Ic} using a large, homogeneous dataset from ZTF, processed and curated within a self-consistent framework, and confirm a color difference that is statistically significant. \rev{Then we examine both extrinsic and intrinsic origins of this difference and argue that it most likely has an intrinsic origin.}

\section{SN data}\label{sec_obs}

We obtained SNe Ib/Ic photometric data from ZTF through TNS and the Weizmann Interactive Supernova Data Repository (\mbox{WISeREP}\footnote{\url{https://www.wiserep.org/}}, \citealt{Wiserep}). We selected objects for which there is at least one available spectrum with SNe Ib/Ic classifications (see Appendix~\ref{app_class} for \rev{the effects of} potential misclassification). A total of 338 SNe Ib/Ic, including the Ibn and Icn subclasses, were obtained as of 18/11/2024.

We used the ZTF Forced Photometry Service \citep{Masci2019, Masci2023}. We obtained light curves from the start of ZTF to the reference date above. The ZTF forced photometry could not be processed for 109 out of 338 SNe Ib/Ic due to an insufficient number of observations to define a template or baseline flux level for the photometry, and these objects were excluded from the analysis.
A further 23 objects were filtered out because they lacked reliable detections either before or after the peak of the light curve.
The data for the remaining 206 objects were then processed according to the recipes described in the manual \citep{Masci2023}. These included image quality cuts, baseline correction of flux and magnitude estimates, and flux-uncertainty validation. We performed this for the data available from each ZTF CCD per SN and combined them into a single $g$- and $r$-band light curve for each object. 

To obtain reliable $\gr$ values, we selected SNe Ib/Ic whose $g$- and $r$-band light curves are well covered near the time of $r$-band maximum. We only used data within $\pm10$ days with respect to the $r$-band maximum and required at least one pre-$r$-maximum and one post-$r$-maximum data point in both bands. This restriction excluded the cases in which the ZTF light curves monotonically rise or decrease and therefore do not capture the $r$-band maximum, removing further 81 objects. Since ZTF photometric data are not evenly sampled and have different cadences in the $g$- and $r$-bands, we used Gaussian process regression to interpolate light curves over the time interval where both bands have data points (see \Fig{fig_lc}). Our final sample consists of 125 SNe Ib/Ic, including 70 SNe Ib and 55 SNe Ic. This sample is approximately four times larger than the sample of JYB23, which had 33 SNe Ib/Ic (16 SNe Ib and 17 SNe Ic). 

The peak colors, $\gr$, were calculated from the interpolated light curves as the difference between the magnitudes of the $g$- and $r$-bands at the time of the $r$-band maximum. \rev{The colors were corrected for Milky Way reddening \citep{Schlafly2011}, but not for host galaxy reddening (see Sect.~\ref{sec_red}). The colors were then $K$-corrected using \texttt{SNcosmo} \citep[][see Appendix~\ref{app_data}]{Barbary2025}.} The uncertainties in $\gr$ were estimated by propagating the uncertainties of the Gaussian process in both bands at the time. The estimated values for our final sample are listed in \rev{Table~\ref{tab_sn}}.

\section{Results} \label{sec_res}

\begin{figure*}
    \sidecaption
	\includegraphics[width=12cm]{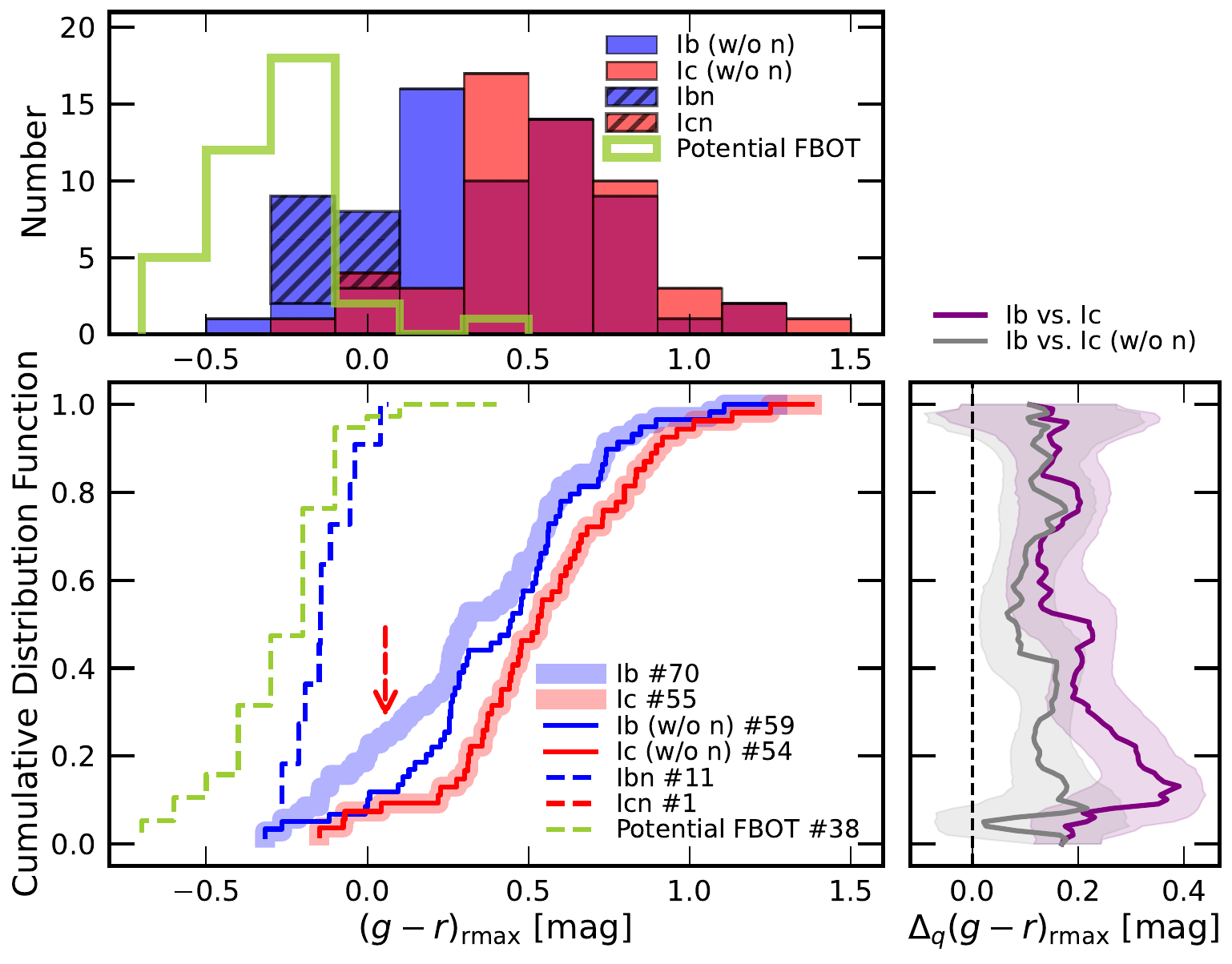}
	\caption{Histograms \rev{(upper left)} and cumulative distributions \rev{(lower left)} of $g-r$ color at the time of $r$-band maximum, $\gr$, for SNe Ib and Ic, \rev{those} without narrow lines, SNe Ibn and Icn, and (potential) FBOTs from \citet{Ho2023}. The contributions from SNe Ibn and Icn are marked differently as “//” hatching \rev{(upper left)} and dashed lines \rev{(lower left). The lower right panel shows differences between the SNe Ib and Ic cumulative distributions as a function of a quantile ($\Delta_q \gr$), for the full sample and for the subsample without narrow lines. Positive values mean that SNe Ib are bluer than SNe Ic. Solid lines show the nominal differences, and shaded regions indicate the 1-$\sigma$ bootstrap uncertainty. The vertical dashed line marks zero difference.}
    }
    \label{fig_cdf}
\end{figure*}

The histograms and cumulative distribution functions in \Fig{fig_cdf} show a clear systematic difference between the two SN subtypes: SNe Ib are bluer than SNe Ic. 
\rev{The mean color difference between SNe Ib and Ic, excluding SNe Ibn and Icn, is $\gap = 0.13 \pm 0.06$ mag $(2.1\sigma)$, where the uncertainty denotes the standard error of the mean.} A Kolmogorov–Smirnov (K-S) test over the two distributions yields a $p$-value of \rev{0.03}, and an Anderson-Darling (A-D) test yields a $p$-value of \rev{0.03, indicating} that the two distributions are unlikely to be drawn from the same parent distribution at a statistically significant level \rev{($p<0.05$)}. \rev{The quantile difference, $\Delta_q \gr$, lies within 0.09-0.16 mag over the bulk of the distribution (gray line in the lower right panel of \Fig{fig_cdf}).} If SNe Ibn and Icn are included, the color difference becomes even larger \rev{to $\gap = 0.20 \pm 0.06$ mag $(3.4\sigma)$, with K-S and A-D $p$-values of 0.001 and 0.001.}

In addition to the color difference between SNe Ib and Ic, we also find distinctly different color distributions between SNe Ibn/Icn and SNe Ib/Ic without narrow lines (\Fig{fig_cdf}). The SNe Ibn/Icn are systematically bluer, exclusively populating the range $\gr \lesssim 0.1$ mag, which produces a noticeable bump in that range in the histograms. This indicates that SNe Ibn/Icn differ from SNe Ib/Ic without narrow lines not only spectroscopically but also photometrically. This is consistent with the findings of \citet{Hosseinzadeh2017}, in which they show that all SNe Ibn spectra in their sample show bluer continua than other \rev{hydrogen-poor} SN \rev{sub}types. 
\rev{Notably, hydrogen-rich SNe also show a similar connection between early narrow line emission and bluer colors  \citep{JacobsonGalan2024}.} 
Interestingly, the color distribution of SNe Ibn/Icn shows a significant overlap with that of potential fast blue optical transients (FBOTs) identified by \citet{Ho2023}. The \rev{mean} color of potential FBOTs is \rev{$\gr = -0.25 \pm 0.03$} mag, which is close to SNe Ibn/Icn in our sample with \rev{$-0.10\pm0.03$} mag\rev{, and these} are significantly bluer than SNe Ib/Ic without narrow lines \rev{with $0.49\pm0.03$} mag. 

\section{Discussion}\label{sec_dis}

\subsection{Host galaxy reddening}\label{sec_red}

If SNe Ic experience stronger reddening than SNe Ib, this can contribute to their redder color. 
\rev{These two subtypes are known to arise from different environments: SNe Ic are more closely associated with star-forming regions than SNe Ib -- implying that SNe Ic progenitors are more massive -- and it has also been suggested that they occur in higher-metallicity environments \citep[e.g.,][]{Anderson2012,Sanders2012}. Such environments can be dustier and produce stronger reddening. In this work, we do not intend to come to a conclusion about whether the reddening is the main driver of the color difference or not, but we note two points.}

\rev{First, several studies report only a small difference in host galaxy reddening between SNe Ib and Ic, with $\Delta E(B-V)_\mathrm{host}=-0.06\dots0.05$ mag (SNe Ic$-$SNe Ib; \citealt{Stritzinger2018}, \citealt{Afsariardchi2021}, \citealt{Rodriguez2023}, excluding the outlier \citealt{Prentice2019}; see Appendix~\ref{app_red}), well below the observed color difference of 0.13 mag in our sample.} \rev{Moreover,} the reddening corrected or minimally reddened samples from JYB23, \citet{Stritzinger2018}, and \citet{Rodriguez2023} \rev{consistently show a positive color difference of 0.10-0.21 mag (Appendix~\ref{app_com}).} These results indicate that the difference does not solely come from the reddening but likely has intrinsic origins. For 25 SNe in our sample \rev{with} host galaxy reddening estimates, correcting for host galaxy reddening yields a \rev{mean} difference of \rev{$\Delta \gr=0.06 \pm 0.07$ mag. This small value most likely reflects a statistical fluctuation in the reddening estimates of this small subsample, rather than a genuine absence of the difference (Appendix~\ref{app_red}). Nonetheless, the host galaxy reddening properties remain poorly constrained, and conclusions based on exiting reddening estimates should be interpreted with caution.}

\rev{Second, even if the observed color difference is solely due to the host galaxy reddening arising from different dust properties in the SN environments, this would still imply that SNe Ic progenitors are more stripped than SNe Ib progenitors (see Sect.~\ref{sec_presn}). Both a higher progenitor mass and a higher metallicity drive stronger stripping through stronger radiation-driven wind. Thus, in either case, the color difference points to a difference in the degree of stripping between SNe Ib and Ic progenitors.}

\subsection{Pre-SN structures}\label{sec_presn}

If SNe Ib arise from helium-rich progenitors and SNe Ic from helium-poor progenitors, differences in ejecta compositions can naturally give rise to different colors. Since helium has a higher recombination temperature than carbon and/or oxygen, helium-rich ejecta experience a faster drop in opacity. JYB23 show that this makes the photosphere recede inward faster, exposing deeper and hotter layers and exhibiting a bluer color at the optical peak \citep[see also fig. 6 in][]{Woosley2021}. They attribute the bluer colors of SNe Ib in their sample to this effect. 

Recent stellar evolution models that follow the stellar evolution zero-age main sequence to near core-collapse with detailed mass-loss prescriptions and binary mass transfer predict the remaining helium mass in SN progenitors to be $\gtrsim0.2\mso$ \citep[e.g.,][]{Gilkis2025,Ercolino2025,Jin2026}. However, detailed spectral models suggest that the amount of helium that can be hidden spectroscopically may be smaller than this \citep[e.g.,][]{Hachinger2012}, although this depends on the ejecta mass, the radiation field, the evolutionary phase of the SN, and the wavelength \citep[e.g.,][]{Dessart2020,Baal2024,Lu2026}. If differences in progenitor structure are indeed the dominant cause of the observed color difference, it will be interesting to investigate how remaining helium masses correlate with colors. In this context, constraining how much helium can be hidden ``photometrically'' -- not just spectroscopically \citep{Lucy1991} -- may provide a unique probe of the helium content in SN Ic progenitors.

The peak color difference between SNe Ibn/Icn and ordinary SNe Ib/Ic \rev{will} provide an important clue about the nature of SNe Ibn/Icn. 
\rev{Their narrow emission lines indicate an interaction with the CSM, which is likely related to their bluer optical colors (e.g., \citealt{Hillier2019} for SNe II).}
Colors near the peak can therefore carry information about the mass and distribution of the CSM \citep[][]{Khatami2024}, offering constraints on the highly uncertain mass loss and binary interaction processes that occur in the final stages of massive star evolution. Interestingly, we also find that the very blue peak colors of SNe Ibn are distinctly similar to the peak colors of FBOTs (\Fig{fig_cdf}): a broader class of events of an as-of-yet unexplained origin \citep{Drout2014}. This might suggest that similar physical conditions are at play in both SNe Ibn and at least some FBOTs \rev{(Appendix~\ref{app_IbnIcnFBOT})}.

\subsection{Explosion properties}
The observed color difference between SNe Ib and Ic might not arise solely from imprints from long-term stellar evolution prior to core collapse, but also from something occurring during or after core collapse. For example, the mass of the synthesized nickel ($^{56}\rm Ni$), the degree of mixing of nickel driven by hydrodynamic instabilities, and the kinetic energy imparted to the ejecta may differ between SNe Ib and Ic. For a given ejecta mass, a larger nickel mass will lead to a bluer color due to stronger heating (JYB23), but SNe Ic seem to have higher nickel masses than SNe Ib with similar ejecta masses \citep{Afsariardchi2021,Barbarino2021,Rodriguez2023}. Variations in kinetic energy within the typical range of \rev{$1-2 \times 10^{51}\, \rm ergs$} do not significantly affect optical colors at peak, but nickel mixing can strongly impact color due to delayed nickel heating \citep{Yoon2019} and line blanketing (JYB23). In addition, highly asymmetric explosions of SNe Ib/Ic \citep[e.g.,][]{Wheeler2002,Janka2016} can contribute to complex radiation fields that may differ between SNe Ib and Ic and contribute to their color difference.

\section{Final remarks}\label{sec_con}

In this Letter, we show that SNe Ib are systematically bluer than SNe Ic at the optical peak, based on a large and homogeneous sample. While host galaxy reddening could contribute to this difference, the currently available data suggest that it is not the sole contributor, pointing toward an intrinsic origin. The most plausible and natural explanation for this color difference \rev{is a} difference in progenitor structure -- helium-rich progenitors for SNe Ib and helium-poor progenitors for SNe Ic, as shown in JYB23. 
\rev{Even if host galaxy reddening were the dominant contributor, the fact that SNe Ib and Ic experience systematically different reddening from their environments would itself point to differences in mass stripping of their progenitors.}
In either case, the observed color difference demonstrates that the two subtypes differ not only spectroscopically but also photometrically and can serve as a powerful observable to probe the nature of SNe Ib/Ic in large photometric surveys such as LSST \citep{Hambleton2023}. This is important because with dedicated modeling in the future, such a peak color difference could help to quantify the key unknowns, such as the mass \rev{and composition of the remaining envelope of SNe Ib/Ic} progenitors and explosion properties. 

Different degrees of mass stripping have been inferred in the progenitors of SNe Ib and Ic from several spectroscopic analyses \citep[e.g.,][]{Hachinger2012,Dessart2015}. 
\rev{The peak color difference can serve as} an additional and independent line of evidence that SNe Ic progenitors are not only hydrogen-stripped but also helium-stripped massive stars. Combined with the comparable occurrence rates between SNe Ib and Ic \citep[e.g.,][]{Graur2017}, \rev{this color difference would place}
important constraints on massive star evolution: helium-stripped SN progenitors must be as frequent as those that retain helium. However, the physical mechanisms responsible for such efficient and frequent late-phase mass stripping remain uncertain. In this context, colors of SNe Ib/Ic provide a valuable probe of \rev{this process}, helping build a more comprehensive picture of the mass loss history of massive stars.

\begin{acknowledgements}
We thank Ruggero Valli, Steve Schulze, Stephen Justham, \rev{Avishay Gal-Yam, and Saurabh Jha} for their help and useful comments. SCY was supported by the NRF RS-2024-00356267. IA acknowledges support from the European Research Council (ERC) under the European Union’s Horizon 2020 research and innovation program (grant agreement number 852097), from the Israel Science Foundation (grant number 2752/19), from the United States - Israel Binational Science Foundation (BSF; grant number 2024812), and from the Pazy foundation (grant number 216312). We acknowledge the Transient Name Server (TNS), the official IAU mechanism for reporting new astronomical transients, and the collaboration behind its reporting and classification. SH acknowledges support during this project through ESO SSDF2025. WEK was supported by the National Science Foundation under Grants No. 2311323, 2206523. CRediT author statement. \textbf{HJ}: Conceptualization, Methodology, Investigation, Visualization, Writing- Original draft preparation, Review \& Editing \textbf{SdM}: Supervision, Project administration \textbf{SH}: Data Curation, Investigation \textbf{JK}: Conceptualization, Writing - Original draft preparation, Review \& Editing \textbf{GC}: Investigation, Writing - Original draft preparation, Review \& Editing \textbf{SCY}: Conceptualization, Writing - Review \& Editing \textbf{IA, WK}: Writing - Review \& Editing.
\end{acknowledgements}

\bibliographystyle{aa}
\bibliography{Reference_list}

\begin{appendix}

\section{Potential spectral misclassification}\label{app_class}

SN spectra evolve with time. If spectra are obtained too late, SNe IIb can be misclassified as SNe Ib, and SNe Ib as SNe Ic \citep[e.g.,][]{Milisavljevic2013,Liu2016}. In other words, some SNe Ib in our sample may be in fact SNe IIb, and some SNe Ic may be SNe Ib. However, about 80\% of our SNe Ib/Ic were classified at $<10\,$days with respect to the $r$-band maximum (Table~\ref{tab_sn}), so we do not expect misclassification to significantly affect our result. 

Contamination of the sample by misclassified SNe would act to reduce the observed color difference. The reason is as follows: SNe IIb appear redder than SNe Ib \citep{Stritzinger2018,Rodriguez2023,Khakpash2024}, so if our SNe Ib sample is contaminated by SNe IIb, it would shift the \rev{mean} SNe Ib color redder. Similarly, SNe Ib are bluer than SNe Ic, so if our SNe Ic sample is contaminated by SNe Ib, it would make the \rev{mean} SNe Ic color bluer. Both effects act to reduce the observed color difference. Therefore, our estimated color difference should represent a conservative limit, and the actual color difference between SNe Ib and Ic without any misclassification is likely larger than estimated here.

\section{SN sample data}\label{app_data}

\begin{figure}
  \centering
  \includegraphics[width=0.75\linewidth]{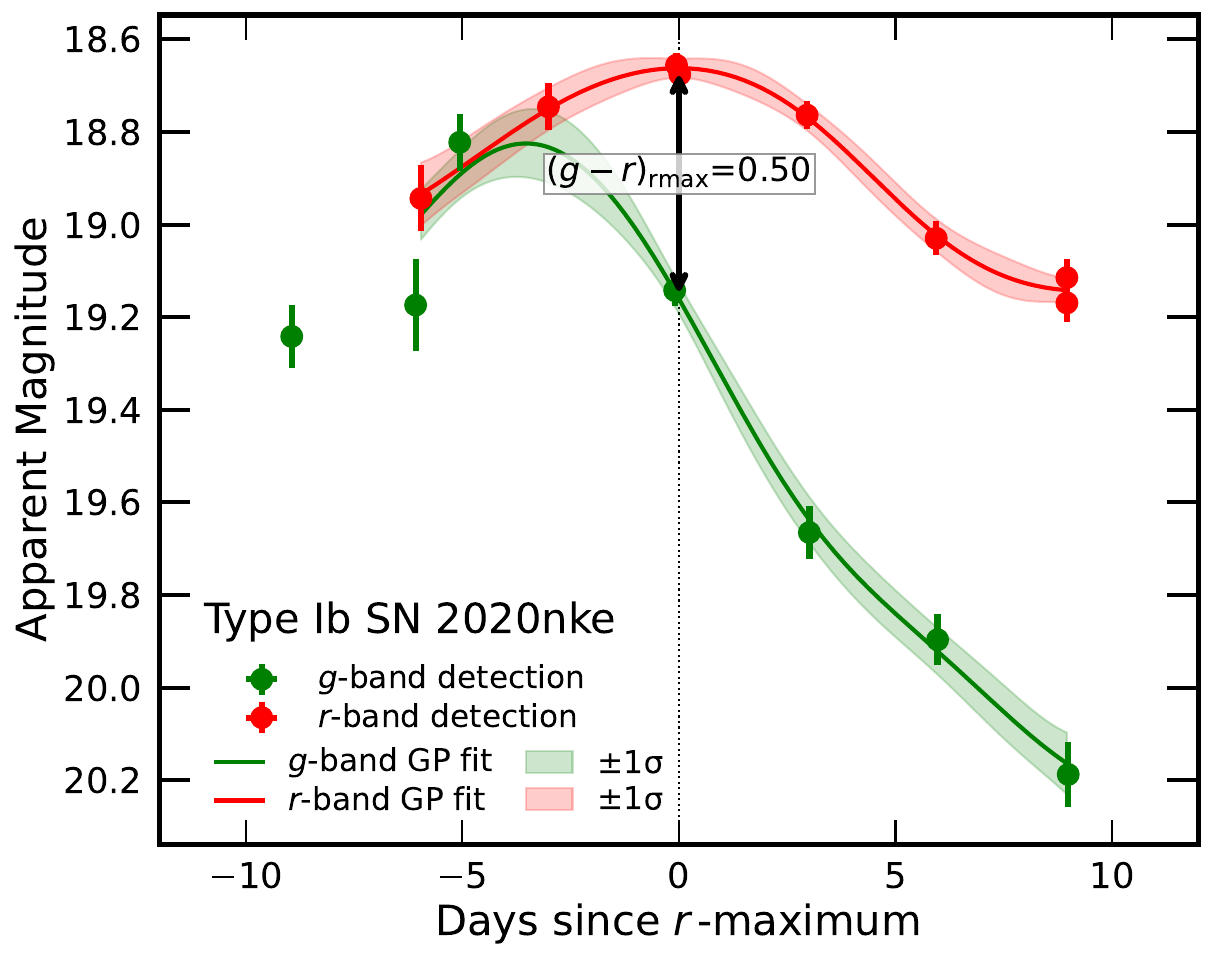}
  \includegraphics[width=0.75\linewidth]{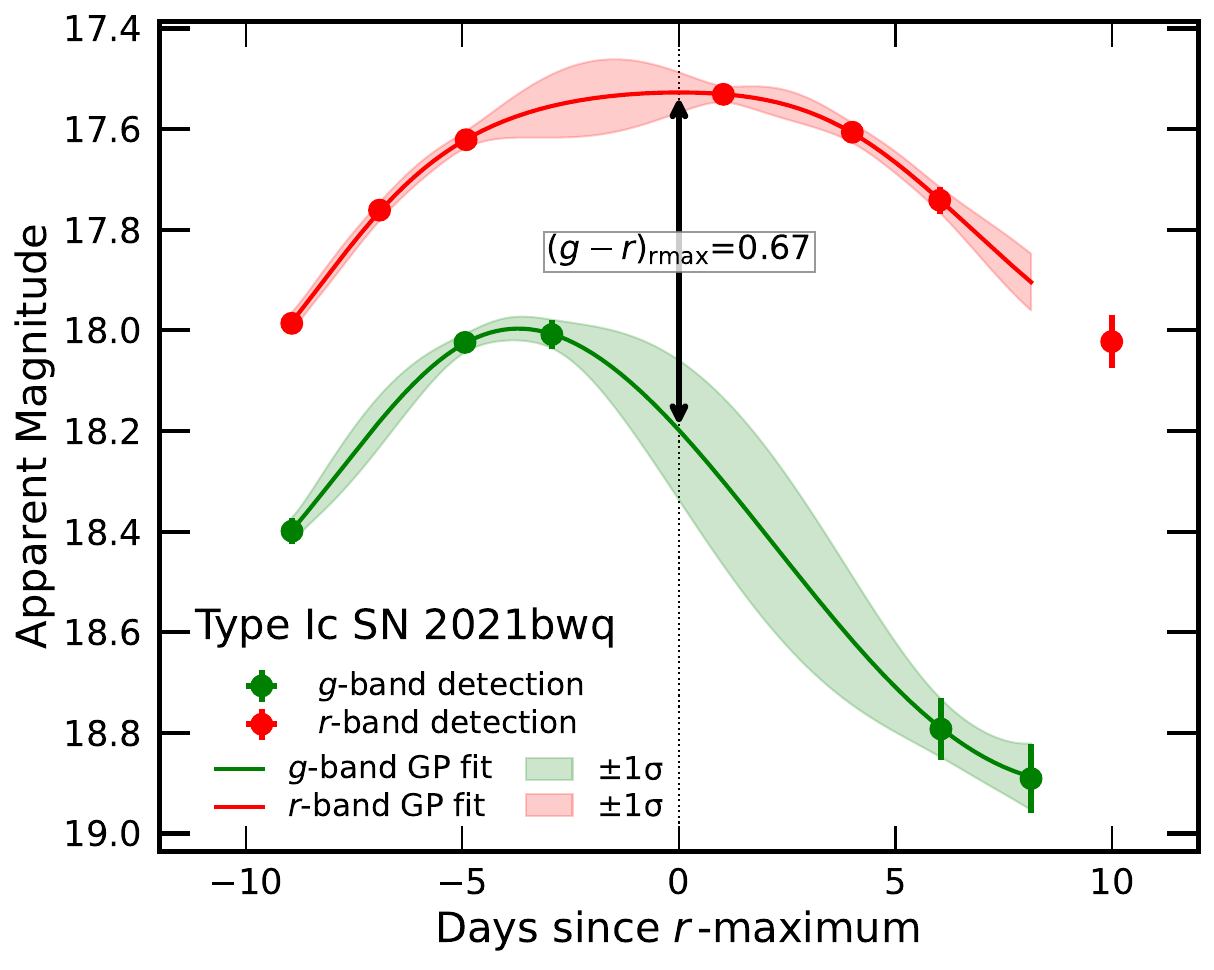}
  \caption{Light curves in $g$- and $r$-bands for one SN Ib \rev{(SN 2020nke)} and one SN Ic (SN 2021bwq), of which $\gr$ values \rev{after Milky Way reddening and $K$-correction} are close to the \rev{mean} values of SNe Ib and Ic \rev{without narrow lines}. The data points show the observation, the solid lines represent the Gaussian Process fits, and the shaded regions indicate the 1-$\sigma$ uncertainties. The black arrows show the determined $\gr$ values. The $x$-axis represents days since the $r$-band maximum, determined from the fitted light curve.}
  \label{fig_lc}
\end{figure}

\begin{figure}
  \centering
  \includegraphics[width=0.8\linewidth]{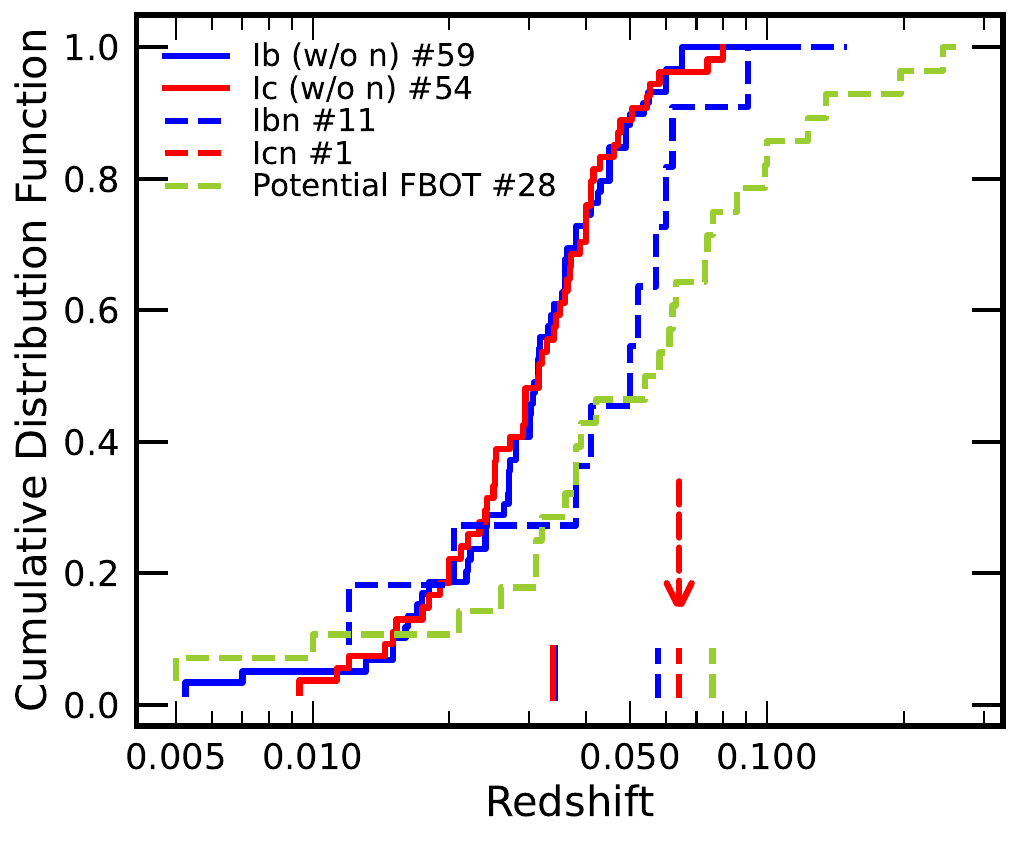}
  \caption{
  \rev{Cumulative redshift distributions of SNe Ib and Ic without narrow lines, SNe Ibn and Icn, and (potential) FBOTs (\citealt{Ho2023}, for which only objects with redshift measurements are shown). Short ticks at the bottom indicate the mean redshift of each kind, with those for SNe Ib and Ic without narrow lines overlapping due to their nearly identical values.}
  }
  \label{fig_redshift}
\end{figure}

Table~\ref{tab_sn} provides relevant photometric parameters derived using our method outlined in Sect.~\ref{sec_obs}. \Figure{fig_lc} shows two representative light curves to showcase how the interpolation of the light curves is done using the Gaussian Process regression. \rev{\Figure{fig_redshift} shows the redshift distributions of different subtypes. SNe Ib and Ic exhibit very similar distributions, which is due to their similar peak luminosities  \citep[e.g.,][]{Taddia2018}. The redshifts of SNe Ibn/Icn and FBOTs are on average higher than ordinary SNe Ib/Ic due to their higher luminosities \citep{Drout2014,Hosseinzadeh2017}, which allow them to be detected at greater distances. We apply $K$-correction using \texttt{SNcosmo} \citep[][]{Barbary2025}, adopting the spectral templates of \citet{Vincenzi2019} for SNe Ib/Ic and a mean spectral template constructed from SNe Ibn spectra from \citet{Hosseinzadeh2017} for SNe Ibn/Icn. As we have only one SN Icn in our sample, we apply the same template as for SNe Ibn. SNe Ib and Ic without narrow lines have nearly identical redshift distributions, so the difference in their $K$-corrections arise from their spectral energy distributions rather than redshift. The corrections shift their color difference $\gap$ by only $0.03$ mag (from $0.16$ mag observed to $0.13$ mag rest-frame) well below the measured color difference.} 

\newpage
\onecolumn

\small

\begin{longtable}{cccccccc}
 \caption{List of SNe Ib/Ic in our sample.}\label{tab_sn}\\
 
\hline \hline 
Name & ${g_\mathrm{rmax}}$ [mag] & ${r_\mathrm{rmax}}$ [mag] & $(g-r)_{r_\mathrm{max}}$ [mag] & ${t_\mathrm{rmax}}$ [MJD] & ${t_\mathrm{spec}}$ [MJD] & ${t_\mathrm{spec}}-{t_\mathrm{rmax}}$ [MJD] & Type \\
\hline 
\endfirsthead

\hline \hline 
Name & ${g_\mathrm{rmax}}$ [mag] & ${r_\mathrm{rmax}}$ [mag] & $(g-r)_{r_\mathrm{max}}$ [mag] & ${t_\mathrm{rmax}}$ [MJD] & ${t_\mathrm{spec}}$ [MJD] & ${t_\mathrm{spec}}-{t_\mathrm{rmax}}$ [days] & Type \\
\hline
\endhead

\hline
\endfoot

\hline
\endlastfoot

SN 2018bvi & 18.01 ± 0.06 & 17.52 ± 0.08 & 0.44 ± 0.10 & 58272.0 & 58260.4 & -11.6 & Ib \\
SN 2018dgx & 18.53 ± 0.01 & 17.87 ± 0.01 & 0.57 ± 0.05 & 58322.0 & 58319.4 & -2.7 & Ic \\
SN 2018fob & 19.59 ± 0.18 & 18.63 ± 0.05 & 0.88 ± 0.19 & 58361.1 & 58351.0 & -10.1 & Ic \\
SN 2018gcj & 19.11 ± 0.02 & 18.59 ± 0.02 & 0.31 ± 0.10 & 58377.9 & 58372.4 & -5.5 & Ic \\
SN 2018gjx & 15.89 ± 0.01 & 15.97 ± 0.00 & -0.15 ± 0.01 & 58379.6 & 58379.3 & -0.3 & Ibn \\
SN 2018ise & 19.25 ± 0.10 & 18.64 ± 0.02 & 0.45 ± 0.13 & 58455.9 & 58479.0 & 23.1 & Ic \\
SN 2018kva & 18.17 ± 0.04 & 17.82 ± 0.02 & 0.23 ± 0.08 & 58486.6 & 58486.4 & -0.2 & Ic \\
SN 2019abb & 18.15 ± 0.17 & 17.67 ± 0.01 & 0.41 ± 0.18 & 58513.6 & 58509.0 & -4.6 & Ic \\
SN 2019agx & 19.28 ± 0.39 & 18.70 ± 0.14 & 0.53 ± 0.42 & 58532.0 & 58526.4 & -5.6 & Ib \\
SN 2019amm & 19.22 ± 0.05 & 18.55 ± 0.04 & 0.60 ± 0.06 & 58579.7 & 58557.4 & -22.3 & Ib \\
SN 2019buy & 18.55 ± 0.03 & 18.19 ± 0.02 & 0.27 ± 0.05 & 58588.8 & 58573.1 & -15.7 & Ib \\
SN 2019dgz & 19.42 ± 0.05 & 18.89 ± 0.04 & 0.48 ± 0.07 & 58595.4 & 58597.3 & 1.9 & Ib \\
SN 2019dld & 18.96 ± 0.05 & 18.19 ± 0.07 & 0.64 ± 0.10 & 58593.8 & 58586.9 & -6.9 & Ic \\
SN 2019dxr & 19.41 ± 0.50 & 18.84 ± 0.03 & 0.48 ± 0.50 & 58606.3 & 58616.2 & 9.8 & Ib \\
SN 2019gqd & 19.21 ± 0.03 & 18.21 ± 0.01 & 0.92 ± 0.07 & 58645.5 & 58642.4 & -3.1 & Ic \\
SN 2019hgp & 18.71 ± 0.02 & 18.68 ± 0.02 & 0.06 ± 0.03 & 58650.8 & 58642.1 & -8.7 & Icn \\
SN 2019hjg & 19.49 ± 0.03 & 18.85 ± 0.02 & 0.53 ± 0.08 & 58659.5 & 58660.4 & 0.9 & Ic \\
SN 2019ilo & 18.84 ± 0.04 & 18.28 ± 0.03 & 0.47 ± 0.08 & 58678.9 & 58669.2 & -9.7 & Ic \\
SN 2019krw & 18.89 ± 0.02 & 18.40 ± 0.02 & 0.36 ± 0.08 & 58679.8 & 58699.3 & 19.5 & Ic \\
SN 2019lsm & 18.08 ± 0.04 & 18.21 ± 0.08 & -0.11 ± 0.09 & 58693.3 & 58695.0 & 1.7 & Ibn \\
SN 2019luy & 19.77 ± 0.15 & 19.12 ± 0.06 & 0.51 ± 0.17 & 58701.2 & 58695.4 & -5.7 & Ib \\
SN 2019oli & 18.14 ± 0.03 & 18.02 ± 0.09 & 0.00 ± 0.10 & 58733.4 & 58748.4 & 15.0 & Ib \\
SN 2019oub & 19.67 ± 0.08 & 18.67 ± 0.07 & 0.62 ± 0.14 & 58776.5 & 58748.5 & -28.0 & Ic \\
SN 2019qvt & 18.84 ± 0.12 & 18.18 ± 0.06 & 0.44 ± 0.14 & 58770.6 & 58767.3 & -3.3 & Ib \\
SN 2019qwo & 19.22 ± 0.12 & 18.89 ± 0.08 & 0.25 ± 0.15 & 58753.6 & 58759.4 & 5.8 & Ib \\
SN 2019rlw & 19.42 ± 0.08 & 18.76 ± 0.06 & 0.60 ± 0.10 & 58767.2 & 58761.4 & -5.8 & Ib \\
SN 2019sjx & 19.70 ± 0.11 & 19.00 ± 0.05 & 0.63 ± 0.13 & 58769.5 & 58792.4 & 22.9 & Ib \\
SN 2019swj & 19.57 ± 0.05 & 19.17 ± 0.06 & 0.25 ± 0.08 & 58785.4 & 58789.0 & 3.6 & Ib \\
SN 2019tls & 17.78 ± 0.15 & 17.36 ± 0.03 & 0.38 ± 0.15 & 58802.3 & 58791.5 & -10.7 & Ib \\
SN 2019txt & 20.91 ± 0.13 & 19.58 ± 0.06 & 1.27 ± 0.14 & 58605.0 & 58616.0 & 11.0 & Ib \\
SN 2019uo & 16.98 ± 0.01 & 17.02 ± 0.01 & -0.04 ± 0.01 & 58508.8 & 58512.7 & 3.9 & Ibn \\
SN 2019wzj & 19.00 ± 0.04 & 17.68 ± 0.03 & 1.13 ± 0.07 & 58847.8 & 58844.9 & -2.9 & Ic \\
SN 2020abdw & 19.21 ± 0.05 & 18.20 ± 0.03 & 0.86 ± 0.08 & 59201.6 & 59191.1 & -10.5 & Ic \\
SN 2020abmp & 20.18 ± 0.11 & 19.24 ± 0.06 & 0.80 ± 0.15 & 59220.0 & 59190.0 & -30.0 & Ic \\
SN 2020abpa & 18.37 ± 0.09 & 17.76 ± 0.08 & 0.51 ± 0.13 & 59192.2 & 59195.2 & 3.0 & Ic \\
SN 2020bcq & 17.05 ± 0.01 & 17.03 ± 0.01 & -0.01 ± 0.01 & 58887.2 & 58881.0 & -6.2 & Ib \\
SN 2020bpf & 18.24 ± 0.05 & 17.85 ± 0.02 & 0.30 ± 0.05 & 58899.4 & 58895.1 & -4.2 & Ib \\
SN 2020cgu & 18.19 ± 0.02 & 17.78 ± 0.01 & 0.30 ± 0.06 & 58899.0 & 58891.1 & -7.8 & Ic \\
SN 2020cln & 20.40 ± 0.06 & 19.57 ± 0.04 & 0.73 ± 0.07 & 58892.1 & 58899.1 & 7.0 & Ib \\
SN 2020clr & 18.84 ± 0.04 & 18.50 ± 0.05 & 0.22 ± 0.10 & 58902.0 & 58893.4 & -8.6 & Ic \\
SN 2020hdh & 17.95 ± 0.06 & 17.37 ± 0.04 & 0.54 ± 0.08 & 58958.0 & 58955.0 & -3.0 & Ic \\
SN 2020hqn & 18.19 ± 0.03 & 17.80 ± 0.05 & 0.31 ± 0.08 & 58970.5 & 58972.3 & 1.8 & Ic \\
SN 2020hvp & 15.68 ± 0.08 & 14.97 ± 0.01 & 0.56 ± 0.09 & 58977.8 & 58962.7 & -15.1 & Ib \\
SN 2020itj & 19.08 ± 0.10 & 18.89 ± 0.09 & 0.13 ± 0.14 & 58981.0 & 58970.0 & -11.0 & Ib \\
SN 2020kba & 18.84 ± 0.08 & 17.91 ± 0.04 & 0.80 ± 0.11 & 59004.7 & 59007.3 & 2.6 & Ic \\
SN 2020ksa & 18.33 ± 0.01 & 18.29 ± 0.02 & 0.01 ± 0.02 & 58996.7 & 59022.8 & 26.1 & Ib \\
SN 2020kxg & 18.36 ± 0.08 & 18.04 ± 0.06 & 0.24 ± 0.10 & 59004.7 & 59012.3 & 7.6 & Ib \\
SN 2020lls & 19.69 ± 0.15 & 18.92 ± 0.07 & 0.63 ± 0.18 & 59009.8 & 59012.0 & 2.2 & Ic \\
SN 2020nac & 18.12 ± 0.17 & 17.36 ± 0.02 & 0.54 ± 0.17 & 59024.6 & 59022.4 & -2.2 & Ib \\
SN 2020nke & 19.16 ± 0.03 & 18.66 ± 0.02 & 0.41 ± 0.04 & 59039.3 & 59032.4 & -6.9 & Ib \\
SN 2020obn & 19.32 ± 0.02 & 18.36 ± 0.01 & 0.90 ± 0.03 & 59052.8 & 59059.2 & 6.4 & Ib \\
SN 2020rlg & 19.17 ± 0.05 & 17.94 ± 0.01 & 1.01 ± 0.07 & 59091.5 & 59091.4 & -0.2 & Ic \\
SN 2020rwz & 19.27 ± 0.11 & 18.57 ± 0.06 & 0.54 ± 0.14 & 59100.0 & 59114.3 & 14.2 & Ic \\
SN 2020sya & 18.74 ± 0.02 & 18.15 ± 0.01 & 0.47 ± 0.06 & 59116.4 & 59112.0 & -4.4 & Ic \\
SN 2020taz & 18.74 ± 0.01 & 18.84 ± 0.03 & -0.15 ± 0.03 & 59113.6 & 59116.5 & 2.9 & Ibn \\
SN 2020twp & 18.93 ± 0.07 & 18.83 ± 0.08 & -0.15 ± 0.15 & 59123.0 & 59136.0 & 13.0 & Ic \\
SN 2020wpq & 19.39 ± 0.13 & 18.83 ± 0.07 & 0.48 ± 0.14 & 59151.6 & 59146.1 & -5.4 & Ib \\
SN 2021aaad & 18.63 ± 0.03 & 17.46 ± 0.02 & 1.06 ± 0.04 & 59497.6 & 59498.1 & 0.4 & Ib \\
SN 2021absd & 19.38 ± 0.05 & 18.44 ± 0.03 & 0.78 ± 0.07 & 59517.8 & 59549.0 & 31.2 & Ib \\
SN 2021bm & 17.85 ± 0.02 & 17.85 ± 0.02 & -0.07 ± 0.06 & 59223.6 & 59232.4 & 8.8 & Ic \\
SN 2021bwq & 18.20 ± 0.14 & 17.53 ± 0.04 & 0.60 ± 0.15 & 59261.3 & 59252.4 & -8.9 & Ic \\
SN 2021do & 17.36 ± 0.02 & 16.50 ± 0.02 & 0.83 ± 0.04 & 59229.4 & 59218.9 & -10.5 & Ic \\
SN 2021dps & 17.92 ± 0.04 & 18.06 ± 0.04 & -0.14 ± 0.06 & 59271.9 & 59284.0 & 12.1 & Ibn \\
SN 2021fxa & 17.72 ± 0.02 & 17.87 ± 0.02 & -0.19 ± 0.03 & 59292.7 & 59309.0 & 16.3 & Ibn \\
SN 2021gvl & 18.88 ± 0.02 & 18.47 ± 0.01 & 0.32 ± 0.07 & 59311.5 & 59316.3 & 4.8 & Ic \\
SN 2021hrj & 18.41 ± 0.02 & 17.45 ± 0.01 & 0.85 ± 0.03 & 59341.3 & 59323.0 & -18.3 & Ib \\
SN 2021jao & 17.95 ± 0.02 & 17.72 ± 0.02 & 0.18 ± 0.03 & 59337.7 & 59324.4 & -13.2 & Ib \\
SN 2021jpk & 18.57 ± 0.02 & 18.63 ± 0.02 & -0.05 ± 0.03 & 59322.6 & 59321.2 & -1.4 & Ibn \\
SN 2021kev & 18.96 ± 0.02 & 18.31 ± 0.02 & 0.58 ± 0.03 & 59334.7 & 59343.4 & 8.7 & Ib \\
SN 2021ofi & 19.04 ± 0.03 & 18.69 ± 0.05 & 0.31 ± 0.06 & 59373.5 & 59377.0 & 3.5 & Ib \\
SN 2021rfv & 19.20 ± 0.05 & 18.28 ± 0.02 & 0.82 ± 0.06 & 59406.7 & 59416.0 & 9.4 & Ib \\
SN 2021rgi & 17.83 ± 0.08 & 17.24 ± 0.02 & 0.53 ± 0.09 & 59413.4 & 59406.1 & -7.3 & Ib \\
SN 2021rjj & 18.17 ± 0.01 & 18.13 ± 0.01 & -0.12 ± 0.04 & 59407.7 & 59426.9 & 19.2 & Ib \\
SN 2021tej & 19.68 ± 0.07 & 19.30 ± 0.06 & 0.26 ± 0.10 & 59410.5 & 59413.6 & 3.1 & Ib \\
SN 2021tiv & 19.44 ± 0.17 & 19.17 ± 0.15 & 0.04 ± 0.25 & 59419.0 & 59427.0 & 8.0 & Ic \\
SN 2021uyc & 17.77 ± 0.03 & 17.31 ± 0.01 & 0.37 ± 0.05 & 59449.1 & 59440.0 & -9.1 & Ic \\
SN 2021zju & 17.49 ± 0.02 & 16.84 ± 0.02 & 0.55 ± 0.03 & 59499.1 & 59485.3 & -13.9 & Ib \\
SN 2022fxc & 19.22 ± 0.08 & 18.43 ± 0.05 & 0.72 ± 0.09 & 59686.0 & 59703.0 & 17.0 & Ib \\
SN 2022gbn & 19.02 ± 0.07 & 18.40 ± 0.05 & 0.56 ± 0.08 & 59685.8 & 59697.0 & 11.2 & Ib \\
SN 2022ibn & 18.74 ± 0.17 & 18.32 ± 0.03 & 0.36 ± 0.18 & 59701.3 & 59712.4 & 11.0 & Ic \\
SN 2022ksf & 18.28 ± 0.03 & 17.95 ± 0.02 & 0.26 ± 0.04 & 59737.5 & 59730.3 & -7.2 & Ib \\
SN 2022oqm & 16.62 ± 0.01 & 16.17 ± 0.01 & 0.42 ± 0.03 & 59784.4 & 59771.3 & -13.1 & Ic \\
SN 2022pda & 17.50 ± 0.05 & 17.46 ± 0.05 & 0.07 ± 0.07 & 59856.8 & 59847.4 & -9.5 & Ibn \\
SN 2022pff & 19.28 ± 0.03 & 19.62 ± 0.04 & -0.26 ± 0.05 & 59763.1 & 59781.3 & 18.2 & Ibn \\
SN 2022pvt & 18.91 ± 0.16 & 18.57 ± 0.09 & 0.28 ± 0.19 & 59806.7 & 59813.5 & 6.9 & Ib \\
SN 2022rsx & 19.45 ± 0.04 & 18.81 ± 0.03 & 0.53 ± 0.08 & 59816.9 & 59819.2 & 2.3 & Ic \\
SN 2022ued & 17.93 ± 0.02 & 18.06 ± 0.02 & -0.32 ± 0.06 & 59869.1 & 59909.3 & 40.2 & Ib \\
SN 2022ujq & 18.86 ± 0.03 & 18.48 ± 0.06 & 0.23 ± 0.07 & 59852.9 & 59860.0 & 7.1 & Ib \\
SN 2022wux & 19.62 ± 0.07 & 18.45 ± 0.03 & 1.11 ± 0.08 & 59871.2 & 59879.3 & 8.0 & Ib \\
SN 2022wwd & 18.37 ± 0.05 & 18.35 ± 0.04 & -0.07 ± 0.08 & 59864.8 & 59866.4 & 1.6 & Ic \\
SN 2022zlz & 18.97 ± 0.08 & 17.84 ± 0.06 & 0.83 ± 0.11 & 59899.1 & 59898.6 & -0.6 & Ic \\
SN 2022zpp & 18.88 ± 0.03 & 18.50 ± 0.02 & 0.31 ± 0.04 & 59900.6 & 59909.3 & 8.7 & Ib \\
SN 2022zzy & 20.68 ± 0.11 & 19.18 ± 0.07 & 1.38 ± 0.14 & 59899.4 & 59900.0 & 0.6 & Ic \\
SN 2023apl & 18.73 ± 0.04 & 18.19 ± 0.03 & 0.44 ± 0.09 & 59987.4 & 59985.4 & -2.0 & Ic \\
SN 2023eeb & 19.55 ± 0.07 & 18.54 ± 0.06 & 0.90 ± 0.12 & 60045.3 & 60046.5 & 1.2 & Ic \\
SN 2023epp & 18.69 ± 0.02 & 18.47 ± 0.02 & 0.11 ± 0.05 & 60056.4 & 60085.3 & 28.9 & Ib \\
SN 2023fbk & 19.23 ± 0.05 & 18.51 ± 0.03 & 0.60 ± 0.09 & 60057.9 & 60085.3 & 27.4 & Ic \\
SN 2023grp & 19.58 ± 0.09 & 18.53 ± 0.05 & 0.96 ± 0.12 & 60094.9 & 60110.3 & 15.4 & Ic \\
SN 2023icl & 18.92 ± 0.04 & 18.59 ± 0.04 & 0.20 ± 0.06 & 60082.0 & 60106.0 & 24.0 & Ib \\
SN 2023jus & 19.87 ± 0.16 & 19.20 ± 0.14 & 0.56 ± 0.22 & 60098.8 & 60103.9 & 5.2 & Ib \\
SN 2023kbd & 19.37 ± 0.03 & 18.59 ± 0.02 & 0.66 ± 0.07 & 60110.5 & 60104.0 & -6.5 & Ic \\
SN 2023kic & 18.51 ± 0.08 & 17.78 ± 0.07 & 0.66 ± 0.11 & 60129.9 & 60143.6 & 13.7 & Ib \\
SN 2023nlj & 19.11 ± 0.04 & 18.33 ± 0.02 & 0.72 ± 0.05 & 60165.7 & 60182.9 & 17.2 & Ib \\
SN 2023oyz & 18.94 ± 0.02 & 18.15 ± 0.02 & 0.68 ± 0.06 & 60182.5 & 60170.9 & -11.6 & Ic \\
SN 2023plg & 17.56 ± 0.07 & 17.22 ± 0.05 & 0.26 ± 0.08 & 60251.6 & 60254.3 & 2.7 & Ib \\
SN 2023qnh & 18.92 ± 0.06 & 18.35 ± 0.17 & 0.45 ± 0.18 & 60193.6 & 60206.2 & 12.6 & Ib \\
SN 2023qxy & 20.48 ± 0.15 & 19.65 ± 0.07 & 0.10 ± 0.20 & 60181.2 & 60194.5 & 13.3 & Ib \\
SN 2023rkg & 18.93 ± 0.28 & 19.17 ± 0.10 & -0.26 ± 0.30 & 60218.9 & 60221.2 & 2.4 & Ib \\
SN 2023sar & 18.29 ± 0.03 & 17.39 ± 0.02 & 0.73 ± 0.05 & 60213.9 & 60217.0 & 3.1 & Ic \\
SN 2023tno & 17.98 ± 0.02 & 17.51 ± 0.01 & 0.39 ± 0.05 & 60213.5 & 60224.5 & 11.1 & Ic \\
SN 2023uqf & 19.90 ± 0.07 & 20.09 ± 0.09 & -0.21 ± 0.11 & 60230.0 & 60229.0 & -1.0 & Ibn \\
SN 2023utx & 20.92 ± 0.10 & 19.51 ± 0.05 & 1.25 ± 0.12 & 60233.4 & 60231.4 & -2.0 & Ic \\
SN 2023vwh & 17.60 ± 0.01 & 17.52 ± 0.01 & 0.04 ± 0.02 & 60252.7 & 60251.5 & -1.2 & Ibn \\
SN 2023wbs & 20.18 ± 0.09 & 19.35 ± 0.06 & 0.74 ± 0.11 & 60255.2 & 60249.6 & -5.7 & Ib \\
SN 2023wbt & 19.17 ± 0.04 & 18.82 ± 0.06 & 0.29 ± 0.08 & 60252.5 & 60248.1 & -4.3 & Ib \\
SN 2023yqp & 19.13 ± 0.05 & 18.35 ± 0.03 & 0.66 ± 0.08 & 60289.1 & 60298.1 & 9.0 & Ic \\
SN 2024fhv & 19.88 ± 0.13 & 19.30 ± 0.10 & 0.44 ± 0.18 & 60410.0 & 60418.0 & 8.0 & Ic \\
SN 2024fia & 19.64 ± 0.10 & 19.12 ± 0.17 & 0.37 ± 0.22 & 60409.3 & 60418.0 & 8.7 & Ic \\
SN 2024gry & 18.35 ± 0.01 & 17.51 ± 0.01 & 0.74 ± 0.03 & 60447.8 & 60432.5 & -15.3 & Ib \\
SN 2024hkc & 20.79 ± 0.09 & 19.88 ± 0.10 & 0.77 ± 0.16 & 60433.0 & 60499.5 & 66.5 & Ic \\
SN 2024jlc & 17.10 ± 0.03 & 16.87 ± 0.01 & 0.15 ± 0.04 & 60505.1 & 60491.0 & -14.2 & Ib \\
SN 2024pul & 18.20 ± 0.03 & 17.63 ± 0.02 & 0.48 ± 0.07 & 60528.6 & 60523.3 & -5.3 & Ic \\
SN 2024pwk & 17.52 ± 0.02 & 17.13 ± 0.04 & 0.28 ± 0.06 & 60527.8 & 60524.5 & -3.3 & Ic \\
SN 2024rbc & 17.45 ± 0.04 & 16.85 ± 0.04 & 0.52 ± 0.06 & 60542.0 & 60535.4 & -6.6 & Ib \\
SN 2024roj & 19.50 ± 0.10 & 18.62 ± 0.05 & 0.73 ± 0.12 & 60543.6 & 60538.3 & -5.3 & Ic \\ \hline

\end{longtable}
\tablefoot{Columns represent SN name, inferred $g$-band and $r-$ band magnitudes, $(g-r)$ color at the time of $r$-band maximum, the date of $r$-band maximum, the date of the first spectrum, its phase relative to the $r$-band maximum, and SN subtype. \rev{The $(g-r)_{r_\mathrm{max}}$ colors are corrected for the Milky Way reddening and $K$-corrected.}}
\twocolumn

\section{\rev{Host galaxy reddening estimates}}\label{app_red}
\begin{table*}[]
\caption{\rev{Summary of mean host galaxy reddening values of SNe Ib and Ic and their differences from recent studies.}}\label{tab_red}
\centering
\renewcommand{\arraystretch}{1.2}
\begin{tabular}{c|c|cc|c}
\hline
\multirow{2}{*}{Reference} & \multirow{2}{*}{Method} & Type Ib & Type Ic & Type Ic $-$ Type Ib \\ \cline{3-5} 
 &  & $E(B-V)_\mathrm{host}$ [mag] & $E(B-V)_\mathrm{host}$ [mag] & $\Delta E(B-V)_\mathrm{host}$ [mag] \\ \hline \hline
\citet{Stritzinger2018}$^{a}$ & Color template & $0.24 \pm 0.09$ (5) & $0.26 \pm 0.06$ (8) & $0.02 \pm 0.11$ (13) \\
\citet{Prentice2019}$^{b}$ & \ion{Na}{I}\,D & $0.03 \pm 0.01$ (6) & $0.70 \pm 0.39$ (3) & $0.67 \pm 0.39$ (9) \\
\citet{Afsariardchi2021}$^{c}$ & Color template & $0.29 \pm 0.07$ (8) & $0.23 \pm 0.09$ (4) & $-0.06 \pm 0.12$ (12) \\
\citet{Rodriguez2023}$^{d}$ & \ion{Na}{I}\,D+Color template & $0.19 \pm 0.02$ (50) & $0.24 \pm 0.03$ (70) & $0.05 \pm 0.04$ (120) \\ \hline
\end{tabular}
\tablefoot{The ``Method'' column indicates how host galaxy reddening was estimated in each reference (see text). The uncertainties represent standard error of the mean.
The numbers in parenthesis indicate the number of objects in each subset. \rev{We compile $E(B-V)_\mathrm{host}$ values as follows.
\tablefoottext{a}{From their table 3.}
\tablefoottext{b}{From their table 1, where values reported as ``negligible'' are treated as zero reddening.}
\tablefoottext{c}{From their table 1.}
\tablefoottext{d}{From their table E7.}}}
\end{table*}

\rev{Two methods have been exploited to estimate host galaxy reddening of SNe. First is using the equivalent width of the \ion{Na}{I}\,D absorption feature in the SN spectra that originates from the host galaxy, and its relation with reddening \citep{Poznanski2012}. However, due to the lack of high resolution spectra for most SNe, this method is subject to large uncertainties. Second is assuming the color evolution of a stripped-envelope SN follows a template for its SN subtype and fitting the reddening required to match the observed color to the template \citep{Stritzinger2018}. However, this can} obscure the intrinsic photometric diversity within each subtype.

\rev{From the literature, we compile host galaxy reddening estimates, $E(B-V)_\mathrm{host}$, for SNe Ib and Ic in Table~\ref{tab_red}. Note that $E(g-r) \approx E(B-V)$ \citep{Schlafly2011}. Three out of four studies report a small difference in the reddening between the two subtypes, with $\Delta E(B-V)_\mathrm{host}$ between $-0.06$ and $0.05$ mag. In contrast, the estimate of \citet{Prentice2019}---the only one based solely on \ion{Na}{I}\,D---is a clear outlier, showing a much larger difference of $0.67\pm0.39$ mag. This is driven by the much higher reddening inferred for their three SNe Ic (0.70 mag) and the much lower reddening inferred for their six SNe Ib (0.03 mag) than others, and its large uncertainty reflects the limited reliability of the \ion{Na}{I}\,D  method. We therefore rely on the other three estimates. Their reddening differences ($<0.05$ mag) are too small to account for the observed color difference (0.13 mag for SNe Ib and Ic without narrow lines). Thus, the observed color difference most likely has an intrinsic origin, although it is not definitive given that the host galaxy reddening remains poorly constrained.
}

\rev{We note that the subsample of \citet{Rodriguez2023} overlapping with ours is not representative of its parent population regarding its reddening properties. The mean host reddening of SNe Ib in this subsample is $E(B-V)_\mathrm{host}=0.13\pm0.05 \; (N=8)$, much lower than SNe Ib in the full sample with $0.19\pm0.02 \; (N=50)$. In contrast, that of SNe Ic in this subsample is $0.24\pm0.04 \; (N=17)$, consistent with SNe Ic in the full sample with $0.24\pm0.03 \; (N=70)$. Thus SNe Ib in this subsample are biased such that they are under-reddened compared to their parent population. Therefore, the small color difference found in this subsample ($\gap=0.06\pm0.07$) is most likely an artifact of small number sampling, rather than evidence that the intrinsic color difference does not exist.}

\section{Comparison with previous studies}\label{app_com}

\begin{table*}[]
\caption{Summary of \rev{mean} $\gr$ values for SNe Ib and Ic and their differences in our sample and others. 
}\label{tab_com}
\centering
\renewcommand{\arraystretch}{1.2}
\begin{tabular}{cc|cc|c}
\hline
\multicolumn{2}{c|}{\multirow{2}{*}{}} & Type Ib & Type Ic & Type Ic $-$ Type Ib \\ \cline{3-5} 
\multicolumn{2}{c|}{} & $\gr$ [mag] & $\gr$ [mag] & $\Delta \gr$ [mag] \\ \hline \hline
\multirow{4}{*}{This work} & Full sample & $0.35 \pm 0.04$ (70) & $0.55 \pm 0.04$ (55) & $0.20 \pm 0.06$ (125) \\
 & SNe Ib/Ic w/o n & $0.43 \pm 0.04$ (59) & $0.56 \pm 0.04$ (54) & $0.13 \pm 0.06$ (113) \\
 & Reddening corrected & $0.24 \pm 0.07$ (8) & $0.29 \pm 0.04$ (17) & $0.06 \pm 0.07$ (25) \\
 & SNe Ibn/Icn & $-0.11 \pm 0.03$ (11) & $0.06$ (1) & $\sim 0.17$ (12) \\ \hline
 &  & $(g-r)_{V_\mathrm{max}}$ [mag] & $(g-r)_{V_\mathrm{max}}$ [mag] & $\Delta (g-r)_{V_\mathrm{max}}$ [mag] \\ \hline
\multirow{3}{*}{JYB23$^{a}$} & Reddening uncorrected & $0.51 \pm 0.08$ (16) & $0.74 \pm 0.09$ (17) & $0.23 \pm 0.12$ (33) \\
 & Reddening corrected & $0.31 \pm 0.04$ (16) & $0.41 \pm 0.03$ (17) & $0.10 \pm 0.05$ (33) \\
 & Minimally reddened & $0.23 \pm 0.04$ (8) & $0.45 \pm 0.05$ (6) & $0.21 \pm 0.06$ (14) \\ \hline
\multicolumn{2}{c|}{\citet{Stritzinger2018}$^{b}$} & $0.07 \pm 0.03$ (3) & $0.29 \pm 0.03$ (3) & $0.21 \pm 0.04$ (6) \\ \hline
\multicolumn{2}{c|}{\citet{Rodriguez2023}$^{c}$} & $0.15 \pm 0.07\, [0.15]$ (25) & $0.35 \pm 0.08\, [0.20]$ (39) & $0.19 \pm 0.11\, [0.25]$ (64) \\ \hline
\multicolumn{2}{c|}{\citet{Khakpash2024}$^{d}$} & $0.40 \pm 0.09\, [0.69]$ (23) & $0.70 \pm 0.14\, [0.72]$ (19) & $0.29 \pm 0.17\, [1.00]$ (42) \\ \hline
\end{tabular}
\tablefoot{For our sample, shown values are under different selection criteria: the full sample, the subsample excluding SNe Ibn/Icn, \rev{that with host galaxy reddening corrected (where available),} and subsample only with SNe Ibn/Icn. For comparison, values from other studies are presented. The uncertainties represent standard error of the mean. \rev{Numbers in parenthesis indicate the number of objects in each subsample. Where available, an alternative uncertainty estimate is presented in square brackets.} We compile $(B-V)_{V\mathrm{max}}$ values, $(B-V)$ color at the time of $V$-band maximum, of SNe Ib/Ic \rev{from other studies}. Since the obtained values are in $B-V$, we convert those values to $g-r$ following \citet{Jester2005} for a rough comparison. The optical maximum typically occurs slightly earlier in $V$-band than $r$-band by $\sim$1\,day in SNe Ib/Ic \citep{Taddia2015}, which should not significantly affect the colors.
\tablefoottext{a}{\rev{Read from their table 1. ``Reddening uncorrected'' sample is only corrected for Milky Way reddening, not for host galaxy reddening.}}
\tablefoottext{b}{Read from their $B-V$ color curves (their fig. 5).}
\tablefoottext{c}{Read from their $B-V$ color curves (their fig. 13). }
\tablefoottext{d}{Read from their fig. 33. \rev{We find a mismatch between their table 6 and fig. 33; we suspect that the $V-U$ column in their table 6 actually represents $V-B$.}}
}
\end{table*}

Table~\ref{tab_com} provides an overview of the peak color values from this work and from other studies, of which samples are independently constructed. \rev{Since the samples in each study are constructed using different selection criteria, we do not compare the absolute colors between studies, but instead the color difference between SNe Ib and Ic, assuming that the two subtypes were selected on the same basis within each study. Nevertheless, some residual systematic effects from sampling differences may still play a role in the comparisons below.}

\rev{First, we consider color differences between SNe Ib and Ic without narrow lines and where host galaxy reddening is not corrected for. The ``SNe Ib/Ic w/o n'' sample of this work ($0.13 \pm 0.06$ mag), the ``Reddening uncorrected'' sample of JYB23 ($0.23 \pm 0.12$ mag), and \citet{Khakpash2024} sample ($0.29 \pm 0.17$ mag) all show positive values, consistent within their uncertainties.}

\rev{Second, we consider samples that are corrected for host galaxy reddening or minimally reddened: the ``Reddening corrected'' sample of this work ($0.06 \pm 0.07$ mag), the ``Reddening corrected'' sample  of JYB23 ($0.10 \pm 0.05$ mag) and their ``Minimally reddened'' sample ($0.21 \pm 0.06$ mag), \citet{Stritzinger2018} sample ($0.21 \pm 0.04$ mag), and \citet{Rodriguez2023} sample ($0.19 \pm 0.11$ mag). All exhibit a positive difference, spanning $0.06-0.21$ mag, although there is a considerable spread across studies, in which selection effects likely played a substantial role. Our ``Reddening corrected'' sample gives the smallest value due to a bias in this small subsample (see Appendix~\ref{app_red}).}

\section{\rev{SNe Ibn/Icn and FBOTs}}\label{app_IbnIcnFBOT}

\rev{The CSM around SNe Ibn/Icn progenitors form through various channels. One promising channel is binary interaction.}
Pre-collapse interactions (e.g., Case BB or BC mass transfer) can produce CSM, leading to SNe Ibn/Icn \citep[e.g.,][]{Ercolino2025}. Constraining these processes through SN color is an exciting prospect because such interactions are thought to play an important role in the formation of gravitational wave sources \citep[e.g.,][]{Laplace2020,RomeroShaw2020,Qin2024} 
and have an impact on the amount of angular momentum in the compact remnant, affecting black hole spins and the prevalence of magnetar-driven explosions \citep[e.g.,][]{Fuller2022, Hu2023}. 

\rev{Several SNe Ibn have been found far from star-forming regions \citep[e.g.,][]{Hosseinzadeh2019}, suggesting that some of them might not have a massive star origin. A possible alternative scenario is that the progenitor is a runaway helium giant that loses material near core collapse, forming CSM responsible for the narrow-line features and blue colors of SNe Ibn (\citealt{Pols1994}; Lee \& Yoon in prep.). This can occur when the initially more massive star in a binary becomes a low-mass stripped helium star through mass transfer, and when its companion explodes first and disrupts the binary, and make the helium star ejected far from its birthplace.}

\rev{Some SNe Ibn/Icn and FBOTs share several observational similarities (e.g., narrow emission lines, similar peak colors), suggesting a possible physical connection. Both may indicate}
the presence of a stripped helium star, low ejecta mass, and CSM interaction. Indeed, about half of the FBOT candidates identified by \citep{Ho2023} in ZTF exhibit spectra that can be classified as core-collapse and SN Ibn is the second most common type. These clues are particularly valuable given that the nature of FBOTs is under active debate. Some FBOTs have been proposed to originate from dynamical encounters between a compact object and a (helium) star \citep[e.g.,][]{Metzger2022,Tsuna2025,Klencki2025}, and such events may spectroscopically appear as SNe Ibn/Icn \citep{Metzger2022}.

\end{appendix}

\end{document}